\documentstyle[aps,prl,multicol]{revtex}
\begin{document}
\draft
 \def\OP{\tensor P}
\def\B.#1{{\bbox{#1}}}
\def\BE {\begin{equation}}
\def\EE {\end{equation}}
\def\BEA {\begin{eqnarray}}
\def\EEA {\end{eqnarray}}
\def\Fbox#1{\vskip0ex\hbox to 8.5cm{\hfil\fboxsep0.3cm\fbox{%
  \parbox{8.0cm}{#1}}\hfil}\vskip0ex}

\title{Invariants for Correlations of Velocity Differences in
  Turbulent Fields} \author {Victor S. L'vov$^{1,2}$, Evgenii
  Podivilov$^{1,2}$ and Itamar Procaccia$^1$}
\address{${^1}$Department of~~Chemical Physics,
 The Weizmann Institute of Science,
  Rehovot 76100, Israel,\\
  ${^2}$Institute of Automation and Electrometry,
 Ac. Sci.\ of Russia,  630090, Novosibirsk, Russia}
 \maketitle
\begin{abstract}
  The phenomenology of the scaling behavior of higher order structure
  functions of velocity differences across a scale $R$ in turbulence
  should be built around the irreducible representations of the
  rotation symmetry group. Every irreducible representation is
  associated with a scalar function of $R$ which may exhibit different
  scaling exponents. The common practice of using moments of
  longitudinal and transverse fluctuations mixes different scalar
  functions and therefore may mix different scaling exponents. It is
  shown explicitly how to extract pure scaling exponents for
  correlations functions of arbitrary orders.
\end{abstract}
\pacs{PACS numbers 47.27.Gs, 47.27.Jv, 05.40.+j}
\begin{multicols}{2}
Traditional measurements of anomalous scaling in turbulence are based
on hot wire technology which yields information about the longitudinal
components of the velocity field $\B.u(\B.r,t)$ \cite{Fri}.
Accordingly it is customary to consider the structure functions of
longitudinal velocity differences:
 \BEA
S_n(R) &=& \left<[\delta u_l(\B.r,\B.R,t)]^n\right> \ , \label{Sn}\\
\delta\B.u(\B.r,\B.R,t)&\equiv& \B.u(\B.r+\B.R,t)-\B.u(\B.r,t)\ ,
 \label{du}\\
 \delta u_\ell(\B.r,\B.R,t) &\equiv&
 \delta\B.u(\B.r,\B.R,t)\cdot\B.R/R \ . 
\label{dul} 
\EEA 
It is well known that these structure functions appear to scale
with scaling exponent $\zeta^\ell_n$ which are anomalous (nonlinear
functions of $n$): 
\BE S_n(R) \sim R^{\zeta^\ell_n} \ . 
\label{scalel}
\EE 
Only recently has it become feasible, due to advances in
experimental technology \cite{96Cam,97Nou,97CB}, and even more so in
computational methods \cite{97Bor,97GLR,97CSNC}, to measure other
components of the velocity field. In particular a number of groups
have focused on the transverse components 
\BE
\delta\B.u_t(\B.r,\B.R,t) \equiv \delta \B.u(\B.r,\B.R,t)-\delta
u_\ell(\B.r,\B.R,t)\B.R/R.  
\EE 
These groups studied the scaling
exponents of the transverse structure functions 
\BE T_n(R)
=\left<|\delta\B.u_t(\B.r,\B.R,t) |^n\right>\sim R^{\zeta^t_n} \ .
\label{Tn}
 \EE
 Two sets of measurements appear to imply that the
scaling exponents $\zeta^\ell_n$ are the same as $\zeta^t_n$ within
experimental uncertainty \cite{96Cam,97Nou}, whereas other numerical
\cite{97Bor,97GLR,97CSNC} and experimental \cite{97CB} studies
indicate the opposite, i.e. that $\zeta^t_n$ are significantly smaller
than $\zeta^\ell_n$ for $n \ge 4$.
 
 The main point of this Letter is to demonstrate that higher order
 structure functions of longitudinal and transverse moments are not
 likely to exhibit clean scaling behavior, since they mix different
 scalar functions of $R$ which may scale with different scaling
 exponents. In experimental and numerical studies in which all the
 components of the velocity field are available it is advisable to
 consider moments that are invariant under rotations \cite{96LPP};
 such invariants are expected to scale with pure scaling exponents
 that can be extracted from the data.
 
 The problem of mixing of different scalar functions does not exist
 for the second and third order moments of the longitudinal and
 transverse components.  It is worthwhile to go in detail through the
 analysis of the 2nd order moment in order to see why the longitudinal
 and transverse components are not a good choice, and why at the end
 it does not matter at this order. In an isotropic homogeneous medium
 without helicity (with inversion symmetry) the relevant symmetry
 group is the rotation group SO(3) whose irreducible representations
 can be expressed using the spherical harmonics $Y_{\ell,m}$. The most
 general form of the second order moment of velocity differences has
 contributions from $\ell=0$ and 2:
 \BEA &&\langle \delta
 u^\alpha(\B.r,\B.R,t)\delta u^\beta(\B.r,\B.R,t)\rangle=
 \delta_{\alpha\beta} a_0(R)\nonumber \\
 &&+\left[\delta_{\alpha\beta}-{3R_\alpha R_\beta\over
     R^2}\right]a_2(R) \ . \label{second} 
\EEA 
The coefficients in
 this expression carry the index $\ell$, multiplying terms that are
 irreducible representations of the rotation group of dimension
 $2\ell+1$.  The dimension of the irreducible representation is the
 number of tensor components that transform to one another upon
 rotation of the system of coordinates. All the tensor components of a
 given irreducible representation with a given value of $\ell$ must
 have the same coefficient which depends only on $R$. On the other
 hand the scalar functions $a_0(R)$ and $a_2(R)$ may have different
 scaling exponents.
 
 Computing now the longitudinal and transverse moments we find
 \BEA
 \langle \delta u_\ell \delta u_\ell\rangle={R^\alpha R^\beta\over
   R^2}
 \langle \delta u^\alpha\delta u^\beta\rangle &=&a_0(R)-2a_2(R),\\
 \langle \delta \B.u_t \cdot\delta \B.u_t\rangle=\langle|\delta
 \B.u|^2\rangle- \langle \delta u_\ell \delta u_\ell\rangle
 &=&2a_0(R)+2a_2(R) . 
 \EEA 
Obviously these moments mix the two scalar
 functions with different weights.  Fortunately the incompressibility
 constraint forces $a_0(R)$ and $a_2(R)$ to have the same scaling
 exponent. We compute 
\BE {\partial\over \partial R^\alpha}\langle
 \delta u^\alpha\delta u^\beta\rangle= {R^\beta\over R}[{da_0\over
   dR}-2 {da_2\over dR}-6{a_2(R)\over R}] =0 \ , \label{incom}
 \EE
 meaning that the two functions must have the same $R$ scaling, and
 therefore also the 2nd order longitudinal and transverse components
 scale with the same exponents.
 
 The purity (and identity) of exponents of longitudinal and transverse
 fluctuations also holds for the third order moments. The most general
 form of the third order tensor $\langle\delta u^\alpha\delta
 u^\beta\delta u^\gamma\rangle$ has contributions from $\ell=1$ and 3:
\begin{eqnarray}
&&\langle\delta u^\alpha\delta u^\beta\delta u^\gamma\rangle
=b_1(R)[\delta_{\alpha\beta}R^\gamma+\delta_{\alpha\gamma}
R^\beta+\delta_{\beta\gamma}R^\alpha] \nonumber \\
&&+b_3(R)[\delta_{\alpha\beta}R^\gamma+\delta_{\alpha\gamma}
R^\beta+\delta_{\beta\gamma}R^\alpha-5R^\alpha R^\beta
 R^\gamma/R^2] \ . \nonumber
\end{eqnarray}
We again have two distinct scalar functions, each multiplying a
rotationally invariant form, and scaling with potentially different
scaling exponents. Nevertheless, the incompressibility constraint
provides one relation among the scalar functions, leaving us with one
unknown.  Kolmogorov showed \cite{41Kol} that the rate of energy
dissipation fixes the value of the remaining unknown. The form of the
3rd order tensor is thus fully determined, and a calculation shows
that $\left<|\delta u_\ell(\B.r,\B.R,t)|^3\right> \sim \left<|\delta
  u_t(\B.r,\B.R,t)|^3\right>\sim R$.

The first nontrivial example is the 4th order tensor $\langle\delta
u^\alpha\delta u^\beta \delta u^\gamma \delta u^\delta\rangle$. The
most general form of this tensor has contributions with $\ell=0,2$ and
4: 
\BEA \langle\delta u^\alpha\delta u^\beta\delta u^\gamma \delta
u^\delta\rangle
&=&c_0(R) D_0^{\alpha\beta\gamma\delta}+
 c_2(R) D_2^{\alpha\beta\gamma\delta}\nonumber\\
&+&c_4(R)D_4^{\alpha\beta\gamma\delta} \ , \label{gen4} \EEA where \FL
\BEA &&D_0^{\alpha\beta\gamma\delta}={1\over
  \sqrt{45}}\big[\delta_{\alpha\beta}\delta_{\gamma\delta}+
\delta_{\alpha\gamma}\delta_{\beta\delta}
+\delta_{\alpha\delta}\delta_{\beta\gamma}\big] \ , \label{d0}\\
&&D_2^{\alpha\beta\gamma\delta}={1\over \sqrt{28} R^2}\Big[R^\alpha
R^\beta\delta_{\gamma\delta} +R^\alpha
R^\gamma\delta_{\beta\delta}+R^\alpha R^\delta\delta_{\beta\gamma}
\nonumber \\&&+R^\beta R^\gamma\delta_{\alpha\delta}+R^\beta
R^\delta\delta_{\alpha\gamma} +R^\gamma
R^\delta\delta_{\alpha\beta}\Big]-\sqrt{{5\over
    7}}D_0^{\alpha\beta\gamma\delta}
\ , \label{d2}\\
&&D_4^{\alpha\beta\gamma\delta}=\sqrt{{35\over 8}}{R^\alpha R^\beta
  R^\gamma R^\delta\over R^4} -\sqrt{{5\over 2}}
D_2^{\alpha\beta\gamma\delta} -\sqrt{{7\over 8}}
D_0^{\alpha\beta\gamma\delta} \ . \label{d4} 
\EEA 
We see that in this
case we have three independent scalar functions of $R$, i.e. $c_0(R),
c_2(R)$ and $c_4(R)$, which in principle may have different scaling
exponents. In this case the incompressibility constraint furnishes no
relation between these functions; the reason is that there exist
contributions in this tensor like $\langle u^\alpha(\B.r,t)
u^\beta(\B.r,t) u^\gamma(\B.r+\B.R,t)u^\delta(\B.r+\B.R,t)\rangle$,
and the divergence of such a contribution (with summation on any
tensor index) is not zero. In fact, incompressibility no longer places
constraints for any of the higher order correlation functions for
similar reasons.  We note that there is no known way to justify why
the three scalar functions should have the same dependence on $R$.  We
can compute now the longitudinal and transverse 4th order moments:
\BEA &&\langle (\delta u_\ell)^4\rangle \equiv {1\over R^4}R^\alpha
R^\beta R^\gamma R^\delta
\langle\delta u^\alpha\delta u^\beta\delta u^\gamma \delta
 u^\delta\rangle\label{4l}\\
&&\langle|\delta \B.u_t|^4\rangle \equiv \langle(\delta \B.u_t\cdot
\delta \B.u_t)^2 \rangle\label{4t} \\&&= (\delta_{\alpha\beta}
-{R^\alpha R^\beta\over R^2})(\delta_{\gamma\delta} -{R^\gamma
  R^\delta\over R^2}) \langle\delta u^\alpha\delta u^\beta\delta
u^\gamma \delta u^\delta\rangle \ . \nonumber \EEA A calculation
yields \BEA \langle (\delta u_\ell)^4\rangle &=& {c_0(R)\over
  \sqrt{5}}+{2c_2(R)\over \sqrt{7}}
+c_4(R)\sqrt{{8\over 35}} \ , \label{reultul}\\
\langle|\delta \B.u_t|^4\rangle &=&{8c_0(R)\over
  3\sqrt{5}}-{8c_2(R)\over 3\sqrt{7}} +c_4(R)\sqrt{{8\over 35}} \ .
\label{resultut} \EEA
 We see that these components mix the three
scalar functions with different coefficients. There are two
possibilities: either all the scalar functions have the same leading
scaling exponent, or they have different scaling exponents. In the
first case it is obvious that the longitudinal and transverse moments
share the same scaling exponents. In the second case, for a
sufficiently long inertial range, and for $R\ll L$ where $L$ is the
outer scale of turbulence, the smallest exponent will dominate the
scaling of both moments.  Asymptotically the two moments are expected
to have the same scaling behavior.  However, if the three functions
have different (leading) exponents, data with limited scaling range
may lead to the erroneous conclusion that these moments have different
scaling exponents. It should be stressed that the amplitudes of the
three scalar functions {\em may be not universal}, and different
experiments may lead to different weights in this mixed
representation. This may lead to a possible confusion or to
conflicting results as seen in refs.\cite{96Cam}-\cite{97CSNC}.

The more rational procedure that presents itself in light of this
discussion is to compute the scaling behavior of the {\em invariant}
scalar functions which are associated with the higher order tensors.
To achieve this we use the orthonormality of the irreducible
representations, and observe that
 \BEA c_0(R) &=&
D_0^{\alpha\beta\gamma\delta}
\langle\delta u^\alpha\delta u^\beta\delta u^\gamma \delta
 u^\delta\rangle \ , \label{good1}\\
c_2(R) &=& D_2^{\alpha\beta\gamma\delta}
\langle\delta u^\alpha\delta u^\beta\delta 
u^\gamma \delta u^\delta\rangle \ ,\label{good2}\\
c_4(R)&=&D_4^{\alpha\beta\gamma\delta}\langle\delta u^\alpha\delta
u^\beta \delta u^\gamma \delta u^\delta\rangle \ .\label{good3}
 \EEA
Using the explicit form of the irreducible representations
(\ref{d0})-(\ref{d2}) we can evaluate these functions and find 
\BEA
c_0(R)&\propto& \langle|\delta \B.u|^4 P_0\left({\delta u_\ell
    \over |\delta \B.u|}\right)\rangle \propto 
\langle|\delta \B.u|^4\rangle \ , \label{c0}\\
c_2(R)&\propto& \langle|\delta \B.u|^4 P_2\left({\delta u_\ell
    \over |\delta \B.u|}\right)\rangle \nonumber\\
&\propto& \langle|\delta\B.u|^2\left[3(\delta u_\ell)^2-
  |\delta \B.u|^2\right]\rangle \ , \label{c2}\\
c_4(R)&\propto& \langle|\delta \B.u|^4 P_4\left({\delta u_\ell
    \over |\delta \B.u|}\right) \rangle\nonumber \\
&\propto&\langle 35 \delta u_\ell^4-30\delta u_\ell^2 |\delta
\B.u|^2+3|\delta \B.u|^4\rangle \ , \label{c4} 
\EEA
 where $P_\ell$ are
the standard Legendre polynomials of order $\ell$.  We see that our
scalar functions can be represented as particular combinations of
transverse and longitudinal fluctuations. With data from a turbulent
field $\B.u(\B.r,t)$ one can compute in this way each of the
independent scalar functions. Plotting them in double logarithmic
plots (to get rid of the nonuniversal amplitudes) one has a good
chance of extracting pure scaling behavior. After doing so one can
return to the analysis of the longitudinal and transverse components
with some understanding of the leading and subleading scaling
exponents, to control the apparent scaling behavior in limited scaling
ranges.

These considerations are readily extended to higher order moments.
The $n$th order tensor of velocity differences across a scalar $R$
will have $n/2+1$ invariant scalar functions for $n$ odd, and
$(n+1)/2$ invariant functions for $n$ even. There is no need to write
down the explicit form of the irreducible representations, since the
structure exhibited by Eqs.(\ref{c0}) -(\ref{c4}) repeats at all
orders. In other words, the independent scalar function $d_\ell^n(R)$
which is the function associated with the irreducible representation
of order $\ell$ in the $n$th rank tensor of velocity differences can
be written in general as
\begin{equation}
d_\ell^n(R) \propto \left< |\delta \B.u|^n P_\ell \left({\delta u_\ell
\over |\delta \B.u|}\right)\right> \ , \quad \ell \le n \ , \label{dl}
\end{equation}
where $\ell$ has the same parity as $n$. Thus by simply examining the
Legendre polynomials in any textbook one can determine the precise
combination of longitudinal and transverse fluctuations that is
expected to scale with a pure exponents for any order $n$.

In conclusion, it appears extremely worthwhile, in light of the growing
abundance of high quality data on full turbulent velocity fields, to 
implement the approach detailed above. Since one confronts limited
scaling ranges in most applications, it is mandatory to attempt to 
separate leading from subleading scaling contributions in order to be
able to make substantial conclusions about the numerical values of
scaling exponents. The procedure outlined above goes some way in
this direction.
\acknowledgments 
This work was supported in part
by the German Israeli Foundation, the US-Israel Bi-National Science
Foundation, the Minerva Center for Nonlinear Physics, and the Naftali
and Anna Backenroth-Bronicki Fund for Research in Chaos and
Complexity.

\end{multicols}
\end{document}